\documentclass{article}
\usepackage{amsfonts}
\usepackage{amsmath}

\input{tcilatex}
\begin{document}

\title{Preparing a (quantum) belief system }
\author{V. I. Danilov\thanks{%
Central Mathematic Economic Institute, Russian Academy of Sciences,
vdanilov43@mail.ru.} and A. Lambert-Mogiliansky\thanks{%
Paris School of Economics, alambert@pse.ens.fr}}
\maketitle
\date{}

\begin{abstract}
In this paper we investigate the potential for persuasion linked to the
quantum indeterminacy of beliefs. We first formulate the persuasion problem
in the context of quantum-like beliefs. We provide an economic example of
belief manipulation that illustrates the setting. We next establish a
theoretical result showing that in the absence of constraints on
measurements, any belief state can be obtained as the result of a suitable
sequence of measurements. We finally discuss the practical significance of
our result in the context of persuasion.

\end{abstract}

\section{Introduction}

The theory of persuasion was initiated by Kamenica and Gentskow \cite{KG11}
and further developed in a variety of directions. The subject matter of the
theory of persuasion is the use of an information structure (or measurement)
that generates new information in order to modify a person's state of
beliefs with the intent of making her act in a specific way. The question of
interest is how much can a person, call him Sender, influence another {one},
call her Receiver, by selecting a suitable measurement and revealing its
outcome.\textbf{\ }An example is in lobbying. A pharmaceutical company
commissions to a scientific laboratory a specific study of a drug impact,
the result of which is delivered to the politician. The question of interest
from a persuasion point of view is what kind of study best serves the
company's interest.

Receiver's decision to act depends on her beliefs about the world. In \cite%
{KG11} and related works the beliefs are given as a probability distribution
over a set of states of the world. A central assumption is that uncertainty
is formulated in the standard classical framework. As a consequence the
updating of Receiver's beliefs follows Bayes' rule.

However as amply documented the functioning of the mind is more complex and
often people do not follow Bayes rule. Cognitive sciences propose
alternatives to Bayesianism. One avenue of research within cognitive
sciences appeals to the formalism of quantum mechanics. A main reason is
that QM has properties that reminds of the paradoxical phenomena exhibited
in human cognition. Quantum cognition has been successful in explaining a
wide variety of behavioral phenomena such as disjunction effect, cognitive
dissonance or preference reversal (see \cite{HBQ-17}, \cite{Bubu12}).
Moreover there exists by now a fully developed decision theory relying on
the principle of quantum cognition (see \cite{danalm10, danalmver}).
Therefore in the following we shall use the Hilbert space model to represent
the belief of an individual and capture the impact of new information on
those beliefs. Clearly, the mind is likely to be even more complex than a
quantum system but our view is that the quantum cognitive approach already
delivers interesting new insights in particular with respect to persuasion.

In quantum cognition, the system of interest is the decision-maker's mental
representation of the world. It is represented by a \textit{cognitive state.}
This representation of the world is modelled as a quantum-like system so the
decision relevant uncertainty is of non-classical (quantum) nature. As
argued in (\cite{Dualm16}) this modeling approach allows capturing
widespread cognitive difficulties that people exhibit when constructing a
mental representation of a `complex' alternative (cf \cite{Druckman07}). The
key quantum property that we use is that some characteristics (cf.
properties) of a complex mental object may be "Bohr complementary" that is
incompatible in the decision-maker's mind: they cannot have a definite value
simultaneously. A central implication is that measurements (new information)
modifies the cognitive state in a non-Bayesian well-defined manner.

It turns out that persuasion - measurement operations aimed at moving a
cognitive system into a specific state - has a close counter-part in Physics
(see, for example, \cite{FrSch16}\textbf{)}. When doing experiments in
Quantum Mechanics, one often needs to know the state of the particles before
performing operations on them in order to be able to draw conclusions. In
order to determine that state, physicists `prepare' particles in a definite
quantum state, e.g. by means of filtering: for instance they measure the
spin of a set of particles and keep for further operation those that are in
state + while throwing away the others. By doing so they have effectively
created particles with the spin property +.\ The term preparation is rather
broad - it covers any kind of operations that affects the state of a single
system or the composition of a set of systems. One such operation is
measurement. A von Neumann direct measurement prepares a system in the
(pure) state that obtains as the result of the measurement. Generally, a von
Neumann measurement modifies a quantum state according to the von Neumann-L%
\"{u}ders postulate.

As in the classical context our rational Receiver uses new information to
update her beliefs so that choices based on updated preferences are
consistent with ex-ante preferences defined for the condition (event) that
triggered updating.\footnote{%
A rational quantum-like decision-maker is a decision-maker who has
preferences over items (or actions) that she perceives (represents) as
quantum systems. Her preferences over items with uncertain properties (or
actions with uncertain consequences), satisfy a number of axioms similar to
those in the classical case as we show in \cite{danalmver}. These axioms
secure that the preferences can be represented by an expected utility
function.} In \cite{danalmver}, we learned that a dynamically consistent
rational quantum-like decision-maker updates her beliefs according to the
von Neumann-L\"{u}ders postulate. We take this result as starting point to
investigate the potential of manipulation of Sender when facing a rational
quantum-like Receiver. Our central result in Theorem 1 is that, in the
absence of any constraints on measurements, there always exists a sequence
of direct measurements that secures reaching any target state starting from
any initial state. In terms of persuasion, Sender can always persuade
Receiver to believe anything that favors him. This is in sharp contrast with
the classical setting where the expected posterior must be equal to the
prior - a property labeled Bayesian plausibility in \cite{KG11}. Theorem 1
is of course a theoretical result. Achieving the desired belief state may
require a sequence of measurements that is not practically feasible. Such
measurements may be too costly for Sender to undertake and Receiver could
get annoyed. Classical measurements face similar practical constraints. A
distinction however is that any number of classical measurements can always
be performed as a single measurement while this is generally not true in a
quantum situation.

Kamenica and Gentskow motivated their paper by the fact that attempts to
manipulate command a sizable share of our resources in the economy.
Persuasion is at the heart of advertising, courts hearings, lobbying,
financial disclosures and political campaign among other activities. As
suggested by Akerlof and Schiller \cite{akerlof15}, the power of
manipulation seems however much larger and more determinant than what the
classical approach reveals. Our results suggest that quantum cognition may
provide a better model to explain the extent and power of manipulation in
society. Other approaches to influence on decision-making that appeal to the
quantum formalism have been developped (see e.g. \cite{BagKhre17}). An
important distinction however is that in \cite{BagKhre17}, the authors
consider open quantum systems evolving in time. Our research is a step in a
broader project which aims at proposing an alternative to the foundations of
the functioning of free markets in the spirit of Akerlof and Schiller.

\section{The model}

We have two players Receiver (her) and Sender (him). We are interested in
persuasion aimed at influencing Receiver's choice over actions with
uncertain consequences. The consequences of an action depend on the state of
some system (which is often called Nature). Receiver has some a priori
representation of the system that is a belief about the state of Nature.
Sender can provide new information about the state of Nature by selecting
and performing some investigation (measurement). It is assumed that the
outcome of the investigation is public (or reported truthfully). The new
information triggers a revision of Receiver's beliefs and as a consequence
it affects her choice of action. Sender has an interest in Receiver's action
because his own utility depends on the action chosen by Receiver. Sender's
problem amounts to acting on Receiver's beliefs so as to maximize his
expected utility from Receiver's decision, i.e. to persuade her to act in
ways favorable to him. His means of persuasion are information (signal)
structures, we shall also use the term "measurement devices".

We below provide a brief description of the classical setting and thereafter
we develop our argument in the quantum context.

\subsection{The classical setting}

The classical model of persuasion has been well described in Kamenica and
Gentskow (2011), hereafter KG. There is a set $\Omega $ of states of Nature.
For the sake of simplicity we shall assume that the set $\Omega $ is finite.
Receiver's \emph{belief} (or her belief state) is a probability distribution 
$\beta :\Omega \rightarrow \mathbb{R}$, $\beta \left( \omega \right) \geq 0$,%
\textbf{\ }$\sum_{\omega \in \Omega }\beta \left( \omega \right) =1$. The
set $\Delta \left( \Omega \right) $ is the simplex of probability
distributions on $\Omega $.

An \emph{action} is a function $a:\Omega \rightarrow \mathbb{R}$. In a pure
belief state $\beta =\omega $ this action gives (to Receiver) utility $%
a(\omega )$. If our Receiver has a `mixed' belief $\beta $ she expects to
obtain utility $a(\beta )=\sum_{\omega }a(\omega )\beta (\omega )$.

An \textit{information structure} (or a \textit{measurement device}) is a
map $\varphi :\Omega \rightarrow \Delta \left( S\right) ,$ where $S$ is a
set of signals (outcomes) of our measurement device. In a state $\omega \in
\Omega $ this device gives (randomized) signal $\varphi \left( \omega
\right) \in \Delta \left( S\right) .\ $ If we write this more carefully such
a device is given by a family ($f_{s},\ s\in S)$ of functions $f_{s}$ on $%
\Omega ;f_{s}\left( \omega \right) $ gives the probability of obtaining
signal $s$ in state $\omega \in \Omega .$ Of course all the functions $f_{s%
\text{ }}$must be nonnegative and their sum $\sum_{s}f_{s}$ must yield the
unit function $1_{\Omega }$ on $\Omega .$\ 

Assume Sender and Receiver hold common prior $\beta =\left( \beta \left(
\omega \right), \  \omega \in \Omega \right) $.\footnote{%
Common priors is the standard assumption. Allowing for different priors
would require distinguishing between Sender respectively Receiver's belief
where necessary. In addition we may need to assume that Sender knows
Receiver's belief which is also a common assumption.} The probability of
receiving a signal $s$ given the prior $\beta $ is equal to $%
p_{s}=\sum_{\omega }\beta \left( \omega \right) f_{s}\left( \omega \right) .$

More important for us is that our rational classical Receiver uses Bayes'
rule when she receives signal $s$ to update her beliefs, i.e. to form the
posterior $\beta _{s}\in \Delta \left( \Omega \right) ,$ given as $\beta
_{s}(\omega )=f_{s}\left( \omega \right) \beta \left( \omega \right) /p_{s}.$
For Receiver (and for Sender) what is important is the change in the belief
from $\beta $ to $\beta _{s}$ upon receiving signal $s$. Because as she
receives signal $s$ she will choose her optimal action for the updated
beliefs, $a^{\ast }\left( \beta _{s}\right) ,$ and Sender will receive
utility $u\left( a^{\ast }\left( \beta _{s}\right) \right) .$\ On average
when Sender uses such a signal structure (measurement) he receives utility%
\textbf{\ } $\sum p_{s}u\left( a^{\ast }\left( \beta _{s}\right) \right) .$\
And so we can ask which is the best measurement device for Sender? Since $%
\sum_{s}p_{s}\beta _{s}=\beta $, that is the \textit{expected} posterior is
equal to the prior, a property that KG call Bayesian plausibility, the
problem simplifies to finding the signal structure with expected posterior
(equal to the prior) that maximizes the (expected) utility of Sender.

\subsection{The quantum setting}

The description of a quantum system starts with the fixation of a Hilbert
space $H$\ (over the field $\mathbb{R}$\ of real numbers or the field $%
\mathbb{C}$\ of complex number). Physicists usually work with the complex
field $\mathbb{C}$. We, partly for simplicity, shall work with the real
field $\mathbb{R}$, although all goes without changes for the complex case. $%
(\cdot ,\cdot )$\ denotes the scalar product in Hilbert space $H$ (in our
case a finite dimensional space).

We shall be interested not so much in the Hilbert space $H$ as in operators,
that is linear mappings $A:H\rightarrow H$. Such an operator $A$ is \emph{%
Hermitian} (or symmetric) if $(Ax,y)=(x,Ay)$ for all $x,y\in H$. A Hermitian
operator $A$ is \emph{non-negative} if $(Ax,x)\geq 0$ for any $x\in H$. \ 

The notion of trace will be a central instrument in what follows. The trace $%
\mathbf{Tr\ }$of a matrix can be defined as the sum of its diagonal
elements. {It is known} that the trace does not depend on the choice of
basis. With the help of the trace one can introduce the notion of \textit{%
state of a quantum system}. It is defined as a non-negative Hermitian
operator $B$ with trace equal to 1. This notion replaces the classical
concept of probability distribution. The non-negativity of the operator is
analogous to the nonnegativity of a probability measure, and the trace 1 to
the sum of probabilities which equals 1. The set of states is denoted $%
\mathbf{St}=\mathbf{St}(H)$. It is a convex compact subset in the space of
operators. Extreme points of this set are called \emph{pure states} and have
the following form. Let $x\in H$ be a vector in $H$ of the length 1 (that is 
$(x,x)=1$). Define the operator $P_{x}$\ by the formula $P_{x}(y)=(x,y)x$.
Then the operator $P_{x}$ is a pure state, and any pure state has such a
form for an appropriate vector $x$.

General Hermitian operators play the role of classical random variables on $%
\Omega $. In fact for any Hermitian operator $A$ and state $B$ we can define
the `expected value' of $A$ in state $B$ as $\mathbf{Tr}(AB)$. In \cite%
{danalmver} it is shown that a decision(action) subject to non-classical
uncertainty that is an action whose consequence depends on the uncertain
state of a quantum system\textbf{,} can be expressed as a Hermitian
operator. The expected utility of action $a$ represented by operator $A$ in
belief state $B$ is expressed as $\mathbf{Tr}(AB)$ and this number linearly
depends on $B$. In this way, Receiver's preferences over actions are
determined by her belief state $B$ and actions are understood as affine
functions on $\mathbf{St}.$

When it comes to measurement devices (or information structures), we shall
focus on a limited subset of devices referred to as \emph{direct} (or von
Neumann) measurements. These simple devices are sufficient for the purpose
of the present paper. Such a device is given by a family ($P_{s},\ s\in S)$
of projectors with the property $\sum_{s}P_{s}=E$, where $E$ is the identity
operator on $H.$ (A \emph{projector} is an Hermitian operator such that $%
P^{2}=P$. The set $S$ again is understood as the set of signals of the
device.) The probability $p_{s}$ to obtain a signal-outcome $s$ (in a state $%
B\in \mathbf{St}$) is equal to $\mathbf{Tr}\left( P_{s}B\right) $. And the
posterior belief-state is $B_{s}=P_{s}BP_{s}/p_{s}$ (it is easy to check
that it is a state). Here all is standard and simple especially if projector 
$P_{s}$ is one-dimensional (that is a pure state); in this case the
posterior $B_{s}$ is equal to $P_{s}$. If we repeat the measurement we
obtain the same outcome $s$ and the state does not change. This type of
measurement is \emph{repeatable} or "\emph{first kind}". The only thing that
must be underlined is that the \emph{expected posterior} $%
B^{ex}=\sum_{s}p_{s}B_{s}=\sum_{s}P_{s}BP_{s}$ is generally different from
the prior $B$ as is illustrated in the example below. This contrasts with
the classical case and as a consequence the issue of optimality of
measurements for Sender cannot be addressed straightforwardly. Instead, our
objective in this paper is confined to establishing what can be achieved
with a sequence of measurements. \medskip

Quantum measurements be they direct or more general are characterized by a
few distinguishing features. Most importantly, the performance of a
measurement impacts on the state of the system. As a first consequence and
in contrast with the classical case, beliefs do not converge toward complete
information about `a true state' of the system. In the quantum context, the
belief state may be pure, i.e. represent maximal (rather than complete)
information and yet change upon the reception of new information. An
expression of this is that measurements may be incompatible. As a
consequence and in contrast with the classical case, a sequence of direct
measurements cannot generally be merged into a single direct measurement.
Therefore, in this paper we opted for the following approach. On the one
hand we limit ourselves to direct measurements while on the other hand allow
for sequences of direct measurements.

A second consequence of the impact of measurements on the state is the so
called `decoherence' which turns out to be of great value in our context.
Assume that we perform a measurement (for instance a direct one given by the
family of projectors $P_{s}$) but do not learn the result. Such a
measurement can be simply ignored in the classical context. In the quantum
context such a `blind' measurement induces nevertheless a change in the
posterior $B^{\prime }=\sum_{s}P_{s}BP_{s}.$ As we shall see such blind
measurements (clearly useless in a classical context) provide a powerful
mean of changing the beliefs of Receiver. They also have a meaningful
interpretation in the cognitive quantum context.

\section{An illustrative example}

Before introducing our central result we wish to provide an example of
quantum persuasion in a very common context, i.e. when a seller wants to
persuade a potential buyer to purchase an item. So in that situation we
identify Sender with the seller and Receiver with the consumer.

More specifically, our consumer is considering the purchase of a second hand
smartphone at price 30 euros of uncertain value to her, it depends on its
technical quality which may be standard or excellent. She holds subjective
beliefs about the probability that the smartphone is excellent. Based on
those beliefs, she assigns an expected utility value to the smartphone which
determines her decision to buy or not the item.

Let $H$ be a two-dimensional Hilbert space with an orthonormal basis $\left(
e_{1},e_{2}\right) $ and let $(P_{1},P_{2})$ be the corresponding
projectors. Following \cite{danalmver}, the utility of the smartphone is
expressed as operator $A$ which gives a utility equal to 100 in state $%
\left
\vert e_{1}\right \rangle $ (the smartphone is Excellent) and $0$ in $%
\left
\vert e_{2}\right \rangle $ (the smartphone is Standard); in matrix
form this utility can be written as $A=%
\begin{pmatrix}
100 & 0 \\ 
0 & 0%
\end{pmatrix}%
$. Assume further that the utility when not buying the smartphone is $%
\underline{u}=30\ $(she keeps the money). The consumer's (Receiver) decision
is $d\in \left \{ Y,N\right \} $ accept or refuse to buy. The seller
(Sender) receives utility 10 when selling the phone at price 30 whatever its
quality and 0 otherwise.

Assume now that Receivers' belief (her cognitive state) is a pure
(superposed) state represented by vector $b=(1/\sqrt{5},2/\sqrt{5})$ or by
the corresponding projector $B=%
\begin{pmatrix}
1/5 & 2/5 \\ 
2/5 & 4/5%
\end{pmatrix}%
$ in the basis $(e_{1},e_{2})$. Upon a measurement of $B$, we would find
that she assigns probability 1/5 to the state $\left \vert
e_{1}\right
\rangle $ (the smartphone is Excellent) and 4/5 to the state $%
\left \vert e_{2}\right \rangle $ (the smartphone is Standard). Receiver's
expected utility for the smartphone in the belief state $B$ is represented
by the trace of the product of operators $A$ and $B$: 
\begin{equation*}
Eu\left( A;B\right) =\mathbf{Tr}\left( AB\right) =(1/5)\ast 100+(4/5)\ast
0=20<30=\underline{u}.
\end{equation*}%
Given belief $B$ Receiver does not want to buy the smartphone so the seller
earns 0.\medskip

Can Sender persuade Receiver to buy by selecting an appropriate measurement?
We next show that he indeed can induce her to buy with probability 1.
Consider another property (perspective) of the smartphone that we refer to
as Glamour (i.e. whether celebrities have this brand or not). The Glamour
property can be measured with direct von Neumann measurement $(Q_{1},Q_{2})\ 
$with\ two possible outcomes Glamour $\left \vert G\right \rangle $
corresponding to projector $Q_{1}=%
\begin{pmatrix}
1/2 & 1/2 \\ 
1/2 & 1/2%
\end{pmatrix}%
$ and not Glamor $\left \vert NG\right \rangle $ corresponding to $Q_{2}=%
\begin{pmatrix}
1/2 & -1/2 \\ 
-1/2 & 1/2%
\end{pmatrix}%
.$ The Glamour perspective is represented by the basis $\left( \left \vert
G\right \rangle ,\left \vert NG\right \rangle \right) \ $of the state space
(of the mental representation of the smartphone) which is a $45^{\circ }$
rotation of basis $(e_{1},e_{2})$.\footnote{%
The specific relationship between the two properties i.e., a 45$%
{{}^\circ}%
$ rotation can be interpreted as follows. In Receiver's mind (e.g., based on
experience) there is no correlation between the true user value of an item
and its glamour value.} This means that the $\left( \left \vert
e_{1}\right
\rangle ,\left \vert e_{2}\right \rangle \right) $ and $\left(
\left \vert G\right \rangle ,\left \vert NG\right \rangle \right) $ are two
properties (perspectives) that are incompatible in the mind of Receiver. Or
equivalently $\left( P_{1},P_{2}\right) $ and $(Q_{1},Q_{2})$ are
measurements that do not commute with each other. Receiver can think in
terms of either one of the two perspectives but she cannot synthesize
(combine in a\ stable way) pieces of information from the two perspectives.
This is illustrated in figure 1.

\unitlength=1mm \special{em:linewidth 0.4pt} \linethickness{0.4pt} 
\begin{picture}(125.00,63.00)
\put(70.00,10.00){\vector(0,1){40.00}}
\put(70.00,10.00){\vector(1,1){30.00}}
\put(70.00,10.00){\vector(-1,1){30.00}}
\put(70.00,10.00){\vector(2,3){23}}
\put(70.00,10.00){\vector(1,0){40.00}}
\put(113.00,10.00){\makebox(0,0)[cc]{$e_1$}}
\put(70.00,52.00){\makebox(0,0)[cc]{$e_2$}}
\put(107.00,40.00){\makebox(0,0)[cc]{$b'=|G\rangle$}}
\put(33.00,40.00){\makebox(0,0)[cc]{$b''=|NG\rangle$}}
\put(94.00,47.00){\makebox(0,0)[cc]{$b$}}
\bezier{10}(92.50,44.50)(95.00,42.00)(99.00,39.00)
\bezier{40}(92.50,44.50)(78.00,30.00)(65.00,15.00)
\put(98,40.00){\vector(1,-1){1.00}}
\put(66,16.00){\vector(-1,-1){1.0}}
\end{picture}

\begin{center}
Figure 1.
\end{center}

Assume that Sender brings the discussion to the Glamour perspective and
performs the measurement so Receiver learns whether her preferred celebrity
has this smartphone. With some probability $p$ \textbf{(}= 0.9\textbf{)} the
new cognitive state is $B^{\prime }=Q_{1}$ and with the complementary
probability $1-p=.1$ it is $B^{\prime \prime }=Q_{2}$. We note that $%
Eu\left(A;B^{\prime }\right) =\mathbf{Tr}\left( AB^{\prime }\right) =50>30$
and $Eu\left(A;B^{\prime \prime }\right) =\mathbf{Tr}\left( AB^{\prime
\prime }\right) =50>30.$ In both cases Receiver is persuaded to buy and
Sender gets utility 10.

Interestingly the example also illustrates the non-classical phenomenon of
`decoherence'. Namely that the mere fact of performing a measurement \ -
without learning the outcome (a blind measurement) - triggers a significant
change in beliefs and subsequent action. Here the `expected posterior' $%
B^{ex}=pB^{\prime }+(1-p)B^{\prime \prime }=%
\begin{pmatrix}
1/2 & 2/5 \\ 
2/5 & 1/2%
\end{pmatrix}%
\neq B$ in contrast with classical persuasion which is constrained by
Bayesian plausibility meaning that the expected posterior must equal the
prior.

In general, it is not possible to persuade Receiver with probability 1 by
means of a single measurement. However as Theorem 1 below shows it is
theoretically possible to manipulate beliefs with a sequence of appropriate
measurements.

\section{Our central result}

In order to formulate our central result we return in more detail to the
description of direct measurements. A direct measurement device $\mathcal{M}$
will be given by an orthonormal basis $(e_{1},...,e_{n})$, where $n=\dim H$,
and $(e_{i},e_{j})=\delta _{ij}$, and a collection of signals $\left(
s_{1},...,s_{n}\right) $. The vector $e_{i}$ defines the one-dimensional
projector $P_{i};P_{i}\left( x\right) =\left( x,e_{i}\right) e_{i}.$ How
does such a measurement device operate? Upon the reception of signal $s_{i}$
the system transits into state $\left \vert e_{i}\right \rangle $\ $($or\ $%
P_{i})$ with probability $p_{i}=\mathbf{Tr}\left( BP_{i}\right) .$\ If we
represent the initial state $B$ in matrix form in the basis $\left(
e_{1},...,e_{n}\right) ,$\ we have $p_{i}=b_{ii}$ where $b_{ii}$ is the
corresponding diagonal element of the matrix. This is true when the signal
is fully disclosed. If the result is not communicated (a \emph{blind
measurement}) the system transits into the mixed state $\sum p_{i}P_{i}.$
The impact of blind measurements is a distinguishing feature of the quantum
formalism.

A second useful feature of the quantum situation is related to conditional
measurements. Assume that we performed a measurement according to the above
described device (with the signal set $S$). As we receive signal $s$ we may
thereafter perform a new measurement $\mathcal{M}_{s}$ (which can depend on
signal $s$). In the classical context this kind of conditionality does not
play any role because a compound measurement also is a measurement. But in
the quantum context as we shall see further, such a composition is generally
not a direct measurement although it is a measurement in the most general
meaning. We can iterate that procedure conditioning on the result from the
second measurement and so on. As we consider a sequence of measurements, we
assume that Receiver updates her beliefs each time she receives a new piece
of information in the order of reception.

Our main result is that starting from any initial state it is possible to
transit to any state by means of a suitable sequence of conditional
measurements. In term of persuasion and beliefs, it means that Sender can
always persuade Receiver to believe what is most favorable to him in the
sense that Receiver will take the decision (consistent with her preferences)
that is most desirable for Sender independently of her prior. Of course,
this is a purely theoretical result. In practice, it may not be possible to
`play' with Receiver so easily. Measurements are connected with costs,
Receiver may be impatient or get tired of all information etc...\medskip

\textbf{Theorem 1.} \textit{For any prior} $B$, \textit{any target belief
state} $T$, \textit{and any} $\varepsilon >0$, \textit{there exists a
sequence of direct conditional measurements such that with a probability
larger than }$1-\varepsilon $ \textit{the posterior is equal to } $\mathit{T}
$.\medskip

We understand this result as follows: there exists in principle a rather
simple and constructive strategy for Sender to persuade Receiver to believe
anything Sender wants her to believe.\medskip

We first show that the target state can be taken as pure. In fact the final
state $T$ can be represented (by force of the Spectral theorem) as a mixture
of orthogonal pure states, $T=\sum_{i}q_{i}P_{i}$, where $q_{i}\geq 0,$ $%
\sum_{i}q_{i}=1$, $P_{i}$ are projectors on (unit) vector $e_{i}$. Let $P$
be the projector on the following vector $e=\sum_{i}\sqrt{q_{i}}e_{i}$, that
is $P\left( x\right) =\left( x,e\right) e$ for $x\in H$. In the basis $%
(e_{1},...,e_{n})$ the projector $P$ is given by the matrix $(\sqrt{%
q_{i}q_{j}})$\textbf{.}

We assert that if we perform a blind measurement with basis $\left(
e_{1},...,e_{n}\right) ,$ state $P$ transits into state $T$. In fact, the
expected posterior is $P^{ex}=\sum_{i}P_{i}PP_{i}=\sum_{i}q_{i}P_{i}=T.$

Hence, in order to arrive at the state $T$ it is sufficient to arrive at the
pure state $P$. We next show how to transit into any arbitrary pure state $%
P.\ $

Suppose that $P$ is a projector on (unit) vector $\left \vert
e_{1}\right
\rangle .$ We complete it to an orthonormal basis $\left(
e_{1},...,e_{n}\right) $ and as our main measurement $\mathcal{M}$ we take a
non-degenerated (complete) direct measurement in this basis. Whatever the
initial state $B$, after the performance of $\mathcal{M},$ the state
transits into one of the pure states $\left \vert e_{1}\right \rangle
,...,\left \vert e_{n}\right \rangle $ and the signal informs us about which
one. Assume that the signal is $s_{2}$ so the system is now in the state $%
\left \vert e_{2}\right \rangle.$ In that case we construct an auxiliary
direct measurement device $\mathcal{M}_{2}$ with the following orthonormal
basis $\left( e_{1}+e_{2}\right) /\sqrt{2},\  \left( e_{1}-e_{2}\right) /%
\sqrt{2},e_{3},...,e_{n}.$\ If we perform $\mathcal{M}_{2}$ (recalling that
the system is now in $\left \vert e_{2}\right \rangle )$\ the system will
with equal probability transit into $\left \vert \left( e_{1}+e_{2}\right) /%
\sqrt{2}\right \rangle \ $or$\  \left \vert \left( e_{1}-e_{2}\right) /\sqrt{2%
}\right \rangle \ $(we could let $\mathcal{M}_{2}$ be a blind measurement).\
Now we once more apply $\mathcal{M}$\ and with probability 1/2 we obtain the
desired state $\left \vert e_{1}\right \rangle $ (and the undesired state $%
\left \vert e_{2}\right \rangle $ with the same probability). If we obtain $%
\left \vert e_{1}\right \rangle $ we are done. If we obtain $\left \vert
e_{2}\right \rangle $\ we again apply $\mathcal{M}_{2}$ and thereafter $%
\mathcal{M}.$ After $N$ iterations the state will have transited into the
target state $|e_{1}\rangle $ (or $P$) with probability $1-\left( 1/2\right)
^{N}.$

Above we consider the case when the first measurement gave outcome $s_{2}$.
A similar procedure secures the desired state whatever the first outcome $%
s_{i},\ i\neq 1$ . Instead of $\mathcal{M}_{2}$ we use $\mathcal{M}_{i}$
with basis $\left( e_{1}+e_{i}\right) /\sqrt{2},\  \left( e_{1}-e_{i}\right)/%
\sqrt{2},e_{3},...e_{i-1},e_{i+1},...e_{n}$. Which proves the Theorem
1.\medskip

In figure 2 we illustrate the conditional measurement scheme used in the
proof of Theorem 1 for the case when $n=3$.\ $\mathcal{M}_{2(3)}$ represent
blind measurements.

\unitlength=.95mm \special{em:linewidth 0.4pt} \linethickness{0.4pt}

\begin{picture}(147.00,65.00)(-6,0)
\put(70.00,45.00){\circle{7.00}} \put(90.00,35.00){\circle{7.00}}
\put(50.00,35.00){\circle{7.00}} \put(110.00,25.00){\circle{7.00}}
\put(30.00,25.00){\circle{7.00}}
\put(70.00,49.00){\vector(0,1){11.00}}
\put(110.00,29.00){\vector(0,1){12.00}}
\put(30.00,29.00){\vector(0,1){12.00}}
\put(74.00,43.00){\vector(2,-1){12.00}}
\put(94.00,33.00){\vector(2,-1){12.00}}

\put(114.00,23.00){\vector(2,-1){12.00}}
\put(66.00,43.00){\vector(-2,-1){12.00}}
\put(46.00,33.00){\vector(-2,-1){12.00}}
\put(26.00,23.00){\vector(-2,-1){12.00}}
\put(70.00,45.00){\makebox(0,0)[cc]{$\mathcal{M}$}}
\put(110.00,25.00){\makebox(0,0)[cc]{$\mathcal{M}$}}
\put(30.00,25.00){\makebox(0,0)[cc]{$\mathcal{M}$}}
\put(50.00,35.00){\makebox(0,0)[cc]{$\mathcal{M}_3$}}
\put(90.00,35.00){\makebox(0,0)[cc]{$\mathcal{M}_2$}}
\put(143.00,9){\makebox(0,0)[cc]{...}}
\put(-3.00,9){\makebox(0,0)[cc]{...}}
\put(73.00,54.00){\makebox(0,0)[cc]{1}}
\put(33.00,34.00){\makebox(0,0)[cc]{1}}
\put(83.00,41.00){\makebox(0,0)[cc]{2}}
\put(57.00,41.00){\makebox(0,0)[cc]{3}}
\put(113.00,34.00){\makebox(0,0)[cc]{1}}
\put(123.00,21.00){\makebox(0,0)[cc]{2}}
\put(17.00,21.00){\makebox(0,0)[cc]{3}}

\put(130.00,15.00){\circle{7.00}}
\put(130.00,15.00){\makebox(0,0)[cc]{$\mathcal{M}_2$}}

\put(10.00,15.00){\circle{7.00}}
\put(10.00,15.00){\makebox(0,0)[cc]{$\mathcal{M}_3$}}

\put(134.00,13.00){\vector(2,-1){6.00}}
\put(6.00,13.00){\vector(-2,-1){6.00}}
\end{picture}

\begin{center}
Figure 2.
\end{center}

The strategy used in the proof of the theorem has two nice features. First
it does not require knowing the prior $B$ of Receiver. Second it only
appeals to a restricted \textbf{set of types of} measurements, $\mathcal{M}$
and the instrumental $\mathcal{M}_{i}$ ($i=2,...,n$). As earlier mentioned
in practice the optimal strategy will have to take into account costs of
measurements and constraints on the feasible number of iterations.
Therefore, it can be interesting to consider the case when the number of
measurements is limited as we do in \cite{danalm17}.

\section{Discussion}

Manipulation is both an old and a very active field of research in social
sciences (see e.g., \cite{Druckman07}). Recently an approach has received a
lot of attention in the field of economics thanks to the seminal paper by
Kamenica and Gentskow on Bayesian persuasion. The starting point is that one
can manipulate rational people's behavior by acting upon their beliefs. This
is done through the choice of a suitable information structure or
measurement that generates new information. This theory is developed in the
classical uncertainty setting.

An alternative approach to decision-making under uncertainty uses the
quantum formalism to describe (subjective) uncertainty. Relying on the
recent success of quantum cognition in explaining behavioral anomalies, we
have investigated the scope of manipulation when a person's representation
of the world, i.e. her beliefs are represented as a quantum-like system.

Quite remarkably, we establish that a person's beliefs are in theory fully
manipulable. For any target state and any initial state there exists a
sequence of direct measurements such that the target state is reached with a
probability close to one.

In practice, this potential for manipulation is not expected to be realized
because measurements are costly, people have limited patience or because
constructing the appropriate measurements may not be not practically
feasible. However, there are situations where it makes perfect sense. We
provide a simple example showing that under some circumstances a single
measurement in a very common situation can be sufficient to achieve desired
behavior with probability one. The example also illustrates the impact and
power of blind measurements which captures the idea of changing the focus of
a person's mind without bringing any new information. In our example simply
"diverting" Receiver's attention to the Glamour perspective of the
smartphone (that is performing the corresponding blind measurement) modifies
her state of belief so as to make her willing to buy. As noted by Akerlof
and Schiller " just change people's focus and one can change the decisions
they make." \cite[p.173]{akerlof15}.

Our result suggests that the potential for manipulation of human behavior
may indeed be much greater than what is proposed by main stream economic
theory. This finding is in line with recent works in economics that
emphasize the role of manipulation in the functioning of markets.

\pagebreak


\begin{thebibliography}{99}
\bibitem{akerlof15} Akerlof G., and R. Schiller (2015) \emph{Phishing for
Phools - the economics of manipulation and deception}, Princeton University
Press.

\bibitem{BagKhre17} Bagarello F, Basieva I, and A. Khrenikov (2017) \emph{%
Quantum field inspired model of decision-making: Asymptotic stabilisation of
belief state via interaction with surrounding mental environment}, in press
in \textit{Journal of Mathematical Psychology }

\bibitem{Bubu12} Bruza P., and J. Busemeyer (2012) \emph{Quantum Cognition
and Decision-making}, Cambridge University Press.

\bibitem{Druckman07} Chong D. and Druckman James (2007) "Framing Theory" 
\textit{Annual Review of Political Sciences} 10, 103-126.

\bibitem{danalm10} Danilov V. I., and A. Lambert-Mogiliansky (2010).
Expected Utility under Non-classical Uncertainty. \textit{Theory and Decision%
}, 68, 25-47.

\bibitem{danalmver} Danilov V. I., A. Lambert-Mogiliansky, and V.
Vergopoulos (2016) "Dynamic consistency of expected utility under
non-classical (quantum) uncertainty." PSE Working Papers n%
${{}^\circ}$
2016-12 and http://arxiv.org/abs/1708.08244.

\bibitem{danalm17} Danilov V. I., A. Lambert-Mogiliansky (2017), Targeting
in Persuasion Problems. http://arxiv.org/abs/1709.02595.

\bibitem{Dualm16} Dubois F., and A. Lambert-Mogiliansky (2016). Our
(represented) world and quantum-like object, in \emph{Contextuality in
Quantum Physics and Psychology}, ed. Dzafarof et al, World Scientific,
Advanced Series in Mathematical Psychology Vol. 6, 367-387.

\bibitem{FrSch16} Frohlich J., and B. Schubnel (2016) The preparation of
states in Quantum Mechanics. \emph{Journal Math. Phys.} 57, 042101.

\bibitem{HBQ-17} Haven E. and A. Khrenikov editors, (2017), The Palgrave
Handbook of Quantum Models in Social Sciences, Palgrave MacMillan.

\bibitem{KG11} Kamenica E., and M. Gentzkow (2011). Bayesian Persuasion. 
\emph{American Economic Review}, 101(6): 2590-2615.

\bibitem{NCh} Nielsen M.A., and I.L. Chuang (2010) \emph{Quantum Computation
and Quantum Information.} Cambridge Univ. Press, Cambridge.
\end{thebibliography}
\end{document}